\documentclass[conference]{IEEEtran}
\IEEEoverridecommandlockouts

\usepackage{cite}
\usepackage{amsmath,amssymb,amsfonts}
\usepackage{algorithm}
\usepackage{algorithmic}
\usepackage{graphicx}
\usepackage{textcomp}
\usepackage{xcolor}

\begin{document}

\title{Quantum-Channel Matrix Optimization \\for Holevo Bound Enhancement}

\author{\IEEEauthorblockN{Hong Niu$^{a}$, Chau Yuen$^{a}$,~\IEEEmembership{Fellow, IEEE,} Alexei Ashikhmin$^{b}$,~\IEEEmembership{Fellow, IEEE,} and Lajos Hanzo$^{c}$,~\IEEEmembership{Life Fellow, IEEE}}
\IEEEauthorblockA{$^{a}$ The School of Electrical and Electronics Engineering, Nanyang Technological University, Singapore 639798. \\
$^{b}$ The Communications and Statistical Sciences Research Department, Nokia Bell Laboratories, Murray Hill, NJ 07974 USA. \\
$^{c}$ The Department of Electronics and Computer Science, University of Southampton, Southampton SO17 1BJ, U.K. \\(email: hong.niu@ntu.edu.sg; chau.yuen@ntu.edu.sg; alexei.ashikhmin@nokia-bell-labs.com; lh@ecs.soton.ac.uk)}
}
\maketitle

\begin{abstract}
Quantum communication holds the potential to revolutionize information transmission by enabling secure data exchange that exceeds the limits of classical systems. One of the key performance metrics in quantum information theory, namely the Holevo bound, quantifies the amount of classical information that can be transmitted reliably over a quantum channel. However, computing and optimizing the Holevo bound remains a challenging task due to its dependence on both the quantum input ensemble and the quantum channel. In order to maximize the Holevo bound, we propose a unified projected gradient ascent algorithm to optimize the quantum channel given a fixed input ensemble. We provide a detailed complexity analysis for the proposed algorithm. Simulation results demonstrate that the proposed quantum channel optimization yields higher Holevo bounds than input ensemble optimization.
\end{abstract}

\begin{IEEEkeywords}
Quantum communication system, Holevo bound, quantum channel, quantum optimization.
\end{IEEEkeywords}

\IEEEpeerreviewmaketitle

\vspace*{-3mm}
\section{Introduction}

\IEEEPARstart{Q}{uantum} communication \cite{QC1,QC2,QC3} offers advantages over classical systems, particularly in secure information transmission \cite{QC5,QC6,QC7}. Among the various performance metrics in quantum information theory, the Holevo bound \cite{Holevo1,Holevo2,Holevo3} plays a pivotal role in quantifying the maximum classical information reliably transmitted through noisy quantum channels. Optimizing this bound is essential for applications such as quantum-enhanced communication systems \cite{QCS1}, quantum networks \cite{QCN1}, and quantum computing \cite{QCP4}.

However, computing and optimizing the Holevo bound is generally challenging due to its non-additivity and dependence on both the input ensemble and the quantum channel, the latter being constrained by the completely positive trace-preserving (CPTP) condition. Due to this nontrivial structure, closed-form expressions for the Holevo bound are known only for specific classes of channels, such as unital or entanglement-breaking channels \cite{CF1,CF2,CF3}. Thus, the family of general-purpose optimization methods constitutes a widely open area of research.

Recent years have witnessed a surge of interest in optimizing quantum communication systems using numerical and algorithmic techniques \cite{Opt1}. Various studies have explored the optimization of input states \cite{Input1} or measurement strategies \cite{Measurement1} under fixed channel conditions. However, these approaches often assume static or idealized channels, overlooking that many emerging technologies, such as programmable photonic circuits and reconfigurable superconducting systems, enable active channel reconfiguration.

Fundamental results in quantum information theory confirm that arbitrary CPTP maps can be implemented via quantum circuits \cite{ChannelImp4}. For instance, the authors of \cite{ChannelImp1} provide a universal framework based on Stinespring dilation, while the authors of \cite{ChannelImp2} and \cite{ChannelImp3} present more hardware-efficient realizations using CNOT-based constructions.

In practical scenarios, reconfigurable quantum channels offer substantial benefits. They can improve throughput under noise \cite{practice1}, suppress dominant decoherence effects \cite{practice2}, enhance security in quantum key distribution by reducing information leakage \cite{practice3}, and improve gate fidelity in quantum computing processors \cite{practice4}. These capabilities demonstrate that quantum channels can be treated not as fixed constraints, but as tunable resources to be optimized.

Motivated by this context, we aim to maximize the Holevo bound by optimizing the quantum channel with a fixed input ensemble. To tackle this non-convex problem, we propose a projected gradient ascent (GA) algorithm that directly optimizes the Kraus representation of the channel. The method explicitly enforces the physical constraints of the system by ensuring the CPTP constraint at each iteration. Our contributions include:
\begin{enumerate}

  \item \textbf{Problem Formulation:} We formulate the problem of maximizing the Holevo bound via the quantum channel optimization, subject to rigorous physical constraints.

  \item \textbf{Algorithm Design:} We propose projected GA-based algorithms to optimize quantum channel for a given input ensemble. The projection operation maintains the validity of CPTP maps.

  \item \textbf{Complexity Analysis:} We analyze detailed computational complexity characterizations for each component of the proposed algorithm, offering insights into its scalability with respect to the system parameters.

  \item \textbf{Simulation Verification:} Numerical simulations demonstrate that the proposed channel optimization approach significantly improves the Holevo bound compared to input ensemble optimization.

\end{enumerate}


\begin{figure*}
	\centering
	\includegraphics[width=0.8\textwidth]{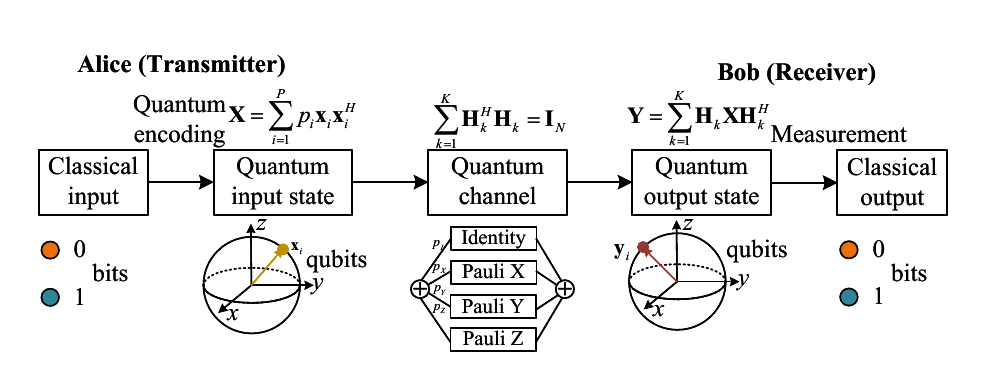}
	\vspace{-3mm}
	\caption{Quantum communication system model.}
    \label{fig_1}
\end{figure*}


\begin{table}[htbp]
\centering
\caption{Notation List}
\vspace{2mm}
\begin{tabular}{l r}
\hline
Symbol & Meaning \\
\hline
${{\mathbf{I}}_S}$ & $S \times S$ identity matrix \\
${\left( \cdot \right)^H}$ & Conjugate transpose \\
$\left\| {\mathbf{x}} \right\|$ & 2-norm of a vector ${\mathbf{x}}$ \\
${x_i}$ & $i$-th element of vector ${\mathbf{x}}$ \\
${\mathrm{Tr}}\left( {\mathbf{X}} \right)$ & Trace of a matrix ${\mathbf{X}}$ \\
$\sum\limits_{i=1}^N x_i$ & Sum of $x_i$ from $i=1$ to $N$ \\
${{\mathbf{X}}^{-1/2}}$ & Matrix inverse square root \\
$\partial f / \partial x$ & Partial derivative of $f$ w.r.t. $x$ \\
$\log x$ & Logarithm of a scalar \\
$\log {\mathbf{X}}$ & Logarithm of a matrix \\
$\mathcal{H}$ & Hilbert space \\
$\mathcal{O}$ & Order of computational complexity \\
$\mathcal{U}(a,b)$ & Uniform distribution over $(a,b)$ \\
$\sim \mathcal{CN}({\mathbf{u}},{\mathbf{\Sigma}})$ & Complex Gaussian distribution \\
$Q(x)$&$\int_x^\infty \frac{1}{\sqrt{2\pi}}e^{-\frac{1}{2}t^2}dt$\\
\hline
\end{tabular}
\label{tbl}
\end{table}

\textit{Notation}: In the following, bold lower-case and upper-case letters denote vectors and matrices, respectively. The notations utilized are summarized in Table I.

\section{System Model}

In this section, we elaborate on the system model of a quantum communication system, including the transmitter (Alice), the receiver (Bob), and a typical quantum channel. As illustrated in Fig. \ref{fig_1}, Alice employs quantum encoding techniques to map classical input bits onto quantum input states, which are subsequently transmitted through a quantum channel \footnote{This may be an optical fiber, a free-space transmission channel, or a circuit-based channel representing the amplitude-damping effect of emitting a photon, and the phase-damping owing to photon scattering.}. Upon reception, Bob performs a measurement on the received quantum output states to retrieve the corresponding classical output bits. Notably, although the overall quantum communication process bears a conceptual resemblance to its classical counterpart, substantial differences exist in the nature of the input and output quantum states, as well as in the properties of the quantum channel. These distinctions will be elaborated in the rest of this section.

\subsection{Quantum Input State}

In quantum communication, the input state is described by a density matrix $\mathbf{X}$, which generalizes pure states to include statistical mixtures \cite{QCP4}. Unlike a pure state vector $\mathbf{x}$, $\mathbf{X}$ captures both quantum superposition and classical uncertainty, in contrast to classical systems where signals are deterministic in amplitude and phase.

Given that a quantum system resides in an $N$-dimensional Hilbert space, the density matrix of a quantum input state can be expressed as
\begin{equation}\label{transmit:1}
{\mathbf{X}} = \sum\limits_{i = 1}^P {{p_i}{{\mathbf{x}}_i}{\mathbf{x}}_i^H} ,
\end{equation}
where ${{\mathbf{x}}_i} = {\left( {{\psi _{i1}},{\psi _{i2}}, \cdots ,{\psi _{iN}}} \right)^T} \in {\mathcal{H}^N}$ denotes a normalized state vector satisfying
\begin{equation}
{\mathbf{x}}_i^H{{\mathbf{x}}_i} = 1,
\end{equation}
${\mathbf{x}}_i^H = \left( {\psi _{i1}^*,\psi _{i2}^*, \cdots ,\psi _{iN}^*} \right) \in {\mathcal{H}^N}$ represents the complex conjugate transpose of ${{\mathbf{x}}_i}$, $P$ is the number of state vectors, $P=1$ corresponds to a pure quantum input state, whereas $P \in {\mathbb{N}^ * }/\left\{ 1 \right\}$ indicates a mixed quantum input state, which is composed of the weighted sum of multiple state vectors, and each state vector ${{\mathbf{x}}_i}$ occurs with a statistical probability ${p_i} \geqslant 0$, with the probabilities satisfying the condition
\begin{equation}\label{con:3}
\sum\limits_{i = 1}^P {{p_i}}  = 1,
\end{equation}
and ${\mathbf{X}} \in {\mathcal{H}^{N \times N}}$ stands for the density matrix that characterizes the statistical quantum state.

By combining (1), (2), and (3), it follows that a physically realizable quantum density matrix satisfies the necessary and sufficient conditions of
\begin{subequations}\label{con:1}
\begin{align}
&{\mathbf{X}} = {{\mathbf{X}}^H}, \label{Za1}\\
&{\mathbf{X}} \succeq 0, \label{Zb1}\\
&{\text{Tr}}\left( {\mathbf{X}} \right) = 1,\label{Zc1}
\end{align}
\end{subequations}
where Eq. (\ref{Za1}) states that ${\mathbf{X}}$ is a Hermitian matrix, ensuring that all its eigenvalues are real, Eq. (\ref{Zb1}) indicates that ${\mathbf{X}}$ is positive semidefinite, implying that all its eigenvalues are non-negative, and Eq. (\ref{Zc1}) corresponds to the normalization condition of a quantum state, representing a quantum system having a total probability of 1.

\subsection{Quantum Channel}

A quantum channel describes how a quantum system evolves. It maps an input quantum state ${\mathbf{X}}$ to an output quantum state ${\mathbf{Y}}$. Mathematically, a quantum channel ${\mathbf{H}}$ is a linear CPTP map characterized by

\begin{equation}\label{channel:1}
{\mathbf{H}}\left( {\mathbf{X}} \right) = \sum\limits_{k = 1}^K {{{\mathbf{H}}_k}{\mathbf{XH}}_k^H,}
\end{equation}
where ${{\mathbf{H}}_k} \in {\mathcal{H}^{M \times N}}, k = 1,2, \cdots ,K$ are Kraus operators describing the effect of the environment or noise process on the state and $K$ represents the number of Kraus operators of the quantum channel, also known as the Kraus rank of the CPTP mapping. The completeness relationship of
\begin{equation}\label{channel:2}
\sum\limits_{k = 1}^K {{\mathbf{H}}_k^H{{\mathbf{H}}_k}}  = {{\mathbf{I}}_N}
\end{equation}
ensures that the map preserves the trace of the output state, i.e., it remains 1.

This model captures both unitary (ideal, noise-free) and non-unitary processes (e.g., decoherence, dissipation, read-out errors), forming the basis for analyzing quantum communication, error correction, and cryptography. It is well-established that any CPTP map via Kraus operators can be physically realized \cite{ChannelImp1,ChannelImp2,ChannelImp3}. For clarity, this paper focuses on ideal quantum channels as in (\ref{channel:1}) and (\ref{channel:2}), without detailing their circuit implementations.

\subsection{Quantum Output State}

When a quantum input state with a density matrix in (\ref{transmit:1}) is transmitted through a quantum channel modeled in (\ref{channel:1}), the quantum output state ${\mathbf{Y}} \in {\mathcal{H}^{M \times M}}$ can be expressed as
\begin{equation}\label{output:1}
{\mathbf{Y}} = {\mathbf{H}}\left( {\mathbf{X}} \right) = {\mathbf{H}}\left( {\sum\limits_{i = 1}^P {{p_i}} {{\mathbf{X}}_i}} \right) = \sum\limits_{i = 1}^P {{p_i}} {\mathbf{H}}\left( {{{\mathbf{X}}_i}} \right),
\end{equation}
where ${{\mathbf{X}}_i} = {{\mathbf{x}}_i}{\mathbf{x}}_i^H$ represents the $i$-th quantum transmission state. Due to the CPTP nature of a quantum channel, the resultant quantum output state ${\mathbf{Y}}$ also satisfies the properties of a physically realizable density matrix shown in (\ref{con:1}).

\subsection{Holevo Bound}

In quantum information theory, the Holevo bound characterizes the maximum amount of classical information that can extracted from a quantum communication process. Specifically, it is defined as the difference between the entropy of the average output state and the average entropy of the individual output states, given by \cite{Holevo1}
\begin{equation}\label{Holevo:1}
C = S\left( {\mathbf{Y}} \right) - \sum\limits_{i = 1}^P {{p_i}} S\left( {{{\mathbf{Y}}_i}} \right),
\end{equation}
where ${{\mathbf{Y}}_i} = \sum\limits_{k = 1}^K {{{\mathbf{H}}_k}{{\mathbf{X}}_i}{\mathbf{H}}_k^H}$ denotes the $i$-th quantum output state generated by applying the quantum channel to the $i$-th quantum input state ${{\mathbf{X}}_i}$. Furthermore, $S\left( {\mathbf{Y}} \right)$ represents the von Neumann entropy of a quantum output state, which serves as a quantum generalization of Shannon entropy. It is calculated as
\begin{equation}\label{Holevo:2}
S\left( {\mathbf{Y}} \right) =  - {\text{Tr}}\left( {{\mathbf{Y}}\log {\mathbf{Y}}} \right) =  - \sum\limits_{m = 1}^M {{\lambda _m}\log {\lambda _m}},
\end{equation}
where $\log {\mathbf{Y}}$ denotes the matrix logarithm (typically computed via the eigenvalue decomposition) and ${\lambda _m}$ are the eigenvalues of ${\mathbf{Y}}$. In this paper, we adopt base-2 logarithms, so the resultant Holevo bound is measured in bits per channel use (bpcu). Furthermore, the quantity $S\left( {{{\mathbf{Y}}_i}} \right)$ represents the entropy of the average output state, quantifying the overall uncertainty across the ensemble, while $\sum\limits_{i = 1}^P {{p_i}} S\left( {{{\mathbf{Y}}_i}} \right)$ reflects the average entropy of the individual output states, quantifying the inherent randomness within each state. Their difference indicates the amount of classical information that can be accessed and reliably extracted from the quantum communication process.

\section{Quantum Channel Optimization and Complexity Analysis}

\subsection{Quantum Channel Optimization}
In this section, we outline a conventional problem of maximizing the Holevo bound by designing a CPTP-constrained quantum channel. Furthermore, we propose an efficient GA algorithm to solve the subproblem of optimizing the quantum channel matrices associated with fixed quantum input states.

In the following, we consider a problem of optimizing the quantum channel matrices ${{{\mathbf{H}}_k}}$, assuming that the quantum input states $\left\{ {{p_i},{{\mathbf{X}}_i}} \right\}$ are fixed. Under this condition, the original problem can be formulated as

\begin{subequations}\label{pf:1}
\begin{align}
\mathop {\max }\limits_{{{\mathbf{H}}_k}} \;\; & C = S\left( {\mathbf{Y}} \right) - \sum\limits_{i = 1}^P {{p_i}} S\left( {{{\mathbf{Y}}_i}} \right), \label{Za}\\
\;\;{\text{s}}{\text{.t}}{\text{.}} \;\; &{{\mathbf{Y}}_i} = \sum\limits_{k = 1}^K {{{\mathbf{H}}_k}{{\mathbf{X}}_i}{\mathbf{H}}_k^H,}   \label{Zb}\\
&\sum\limits_{k = 1}^K {{\mathbf{H}}_k^H{{\mathbf{H}}_k}}  = {{\mathbf{I}}_N}.\label{Zc}
\end{align}
\end{subequations}

Owing to the non-convex nature of the constant modulus constraints in (\ref{Zb}) and (\ref{Zc}), as well as the intricate coupling among optimization variables in (\ref{Za}), obtaining the optimal solution to Problem (\ref{pf:1}) remains a non-trivial task. To address this challenge, closed-form gradient expressions with respect to ${{\mathbf{H}}_k}$ are derived, based on which a GA algorithm is developed for iteratively approaching a high-quality solution.

To tackle Problem (\ref{pf:1}), we develop a GA-based optimization framework. To preserve the CPTP constraint specified in (\ref{Zc}) throughout the iterative process, the algorithm is constructed by analytically calculating the partial derivative with respect to the $k$-th Kraus operator. This formulation inherently maintains the structure required by (\ref{Zc}) during each update. The key procedures of the proposed iterative algorithm are outlined below.

\textit{Step 1: Calculate the partial derivatives}

To proceed with the derivation, we invoke the well-known identity for the gradient of the von Neumann entropy $S\left( {\mathbf{Y}} \right)$ with respect to a Hermitian and positive semidefinite matrix ${\mathbf{Y}}$:
\begin{equation}\label{lemma:1}
\frac{{\partial S\left( {\mathbf{Y}} \right)}}{{\partial {\mathbf{Y}}}} =  - \frac{\partial }{{\partial {\mathbf{Y}}}}{\text{Tr}}\left( {{\mathbf{Y}}\log {\mathbf{Y}}} \right) =  - \log {\mathbf{Y}} - {{\mathbf{I}}_M},
\end{equation}
where ${{\mathbf{I}}_M}$ represents the $M \times M$ identity matrix. This identity follows directly from the known expression for the gradient of the relative entropy $S\left( {{\mathbf{Y}}\left\| {\mathbf{\Sigma }} \right.} \right)  \triangleq  {\text{Tr}}\left( {{\mathbf{Y}}\left( {\log {\mathbf{Y}} - \log {\mathbf{\Sigma }}} \right)} \right)$ given in \cite{Lemma1}, by setting ${\mathbf{\Sigma }} = {\mathbf{I}}$.

Combining (\ref{channel:1}) and (\ref{lemma:1}), the gradient of the von Neumann entropy $S\left( {\mathbf{Y}} \right)$ with respect to ${{{\mathbf{H}}_k}}$ can be derived through the chain rule as
\begin{equation}\label{derivative:1}
\frac{{\partial S\left( {\mathbf{Y}} \right)}}{{\partial {{\mathbf{H}}_k}}} = \frac{{\partial S}}{{\partial {\mathbf{Y}}}} \cdot \frac{{\partial {\mathbf{Y}}}}{{\partial {{\mathbf{H}}_k}}} =  - 2\left( {\log {\mathbf{Y}} + {{\mathbf{I}}_M}} \right){{\mathbf{H}}_k}{\mathbf{X}},
\end{equation}

Similarly, according to the definitions of ${{{\mathbf{Y}}_i}}$ in (\ref{Holevo:1}), the gradient of the von Neumann entropy of the $i$-th output state $S\left( {{{\mathbf{Y}}_i}} \right)$ with respect to ${{{\mathbf{H}}_k}}$ can be expressed as
\begin{equation}\label{derivative:2}
\frac{{\partial S\left( {{{\mathbf{Y}}_i}} \right)}}{{\partial {{\mathbf{H}}_j}}} = \frac{{\partial S}}{{\partial {{\mathbf{Y}}_i}}} \cdot \frac{{\partial {{\mathbf{Y}}_i}}}{{\partial {{\mathbf{H}}_k}}} =  - 2\left( {\log {{\mathbf{Y}}_i} + {{\mathbf{I}}_M}} \right){{\mathbf{H}}_k}{{\mathbf{X}}_i}.
\end{equation}

With the aid of the expressions in (\ref{derivative:1}) and (\ref{derivative:2}), the gradient of the objective function in (\ref{Za}) can be written as
\begin{equation}\label{derivative:3}
\frac{{\partial C}}{{\partial {{\mathbf{H}}_k}}} =  - 2\left( {\log {\mathbf{Y}} + {{\mathbf{I}}_M}} \right){{\mathbf{H}}_k}{\mathbf{X}} + 2\sum\limits_{i = 1}^P {{p_i}} \left( {\log {{\mathbf{Y}}_i} + {{\mathbf{I}}_M}} \right){{\mathbf{H}}_k}{{\mathbf{X}}_i}.
\end{equation}

\textit{Step 2: Update the Kraus operators}

Based on (\ref{derivative:3}), the Kraus operators can be iteratively updated as
\begin{equation}\label{derivative:4}
{{\mathbf{H}}_k} \leftarrow {{\mathbf{H}}_k} + \alpha  \frac{{\partial C}}{{\partial {{\mathbf{H}}_k}}},
\end{equation}
where $\alpha   > 0$ denotes the step size at each iteration.

\textit{Step 3: Normalize the Kraus operators}

In order to ensure that the updated Kraus operators remain compliant with the constraint (\ref{Zc}), a normalization step is applied as
\begin{equation}\label{normalize:1}
{{\mathbf{H}}_k} \leftarrow {{\mathbf{H}}_k}{{\mathbf{G}}^{ - 1/2}},
\end{equation}
where ${\mathbf{G}}$ represents the accumulated Kraus operators defined as
\begin{equation}
{\mathbf{G}} = \sum\limits_{k = 1}^K {{\mathbf{H}}_k^H{{\mathbf{H}}_k}} .
\end{equation}

This normalization ensures that the updated Kraus operators satisfy the CPTP condition in (\ref{Zc}) at each iteration.

By iterating (\ref{derivative:4}) and (\ref{normalize:1}) several times, the objective function $C$ progressively converges. For clarity, Algorithm 1 summarizes the detailed procedures of the proposed GA algorithm.

\begin{algorithm} [h]
        \caption{The Proposed GA Algorithm to solve (\ref{pf:1})}
		\label{alg_approx}
		\begin{algorithmic}[1]
        \REQUIRE
			${{\mathbf{X}}_i},p_i,{{\mathbf{H}}_k}$.
        \REPEAT
        \FOR {$k = 1;k \leqslant K;k +  + $}
           \STATE          Computing the gradient of $C$ via (\ref{derivative:3});
           \STATE          Updating ${{\mathbf{H}}_k}$ by applying (\ref{derivative:4});
           \STATE          Normalizing ${{\mathbf{H}}_k}$ with (\ref{normalize:1});
        \ENDFOR
        \UNTIL The increase of $C$ is lower than a preset threshold or the number of iterations reaches a preset maximum number;
        \ENSURE
			${{\mathbf{H}}_k}$.
		\end{algorithmic}
\end{algorithm}

\subsection{Computational Complexity Analysis}

In this section, we evaluate the computational complexity of the proposed AG scheme in terms of the number of real-valued multiplications. We will show that the closed-form solutions derived exhibits linear complexity with respect to the number of meta-atoms per layer.

For the quantum channel optimization, the computation of the gradient of $C$ w.r.t. ${{\mathbf{H}}_k}$ via (\ref{derivative:3}) has a complexity order of $\mathcal{O}\left[ {\left( {P + 1} \right){M^3} + PM{N^2} + P{M^2}N} \right]$. Moreover, the normalization of ${{\mathbf{H}}_k}$ using (\ref{normalize:1}) entails a complexity of $\mathcal{O}\left( {{N^3} + KM{N^2}} \right)$.

This analysis reveals that the proposed AG algorithm exhibits linear complexity with respect to the number of input states $P$ and the number of Kraus operators $K$, while the complexity scales cubically with respect to the dimensions $M$ and $N$ of the output and input density matrices, respectively.

\section{Simulation Results}

In this section, numerical results are presented to investigate the influence of key system parameters. Performance comparisons are conducted among: (i) Proposed AG algorithm for channel optimization (Algorithm 1); (ii) Conventional input optimization; and (iii) Conventional non-optimized schemes. These comparisons highlight the advantages of the proposed AG algorithms.

\subsection{Simulation Setups}

The simulation parameters are shown in the following. Unless otherwise specified, the dimensions of quantum input and output states are set to $N = 3, M = 4$, and the number of Kraus operators is set to $K = 5$. Meanwhile, the number of input states is set to match the input state dimension, i.e., $P = N$.

Simulation parameters reflect practical quantum communication scenarios. The input dimension is set as $N=3$ (single qutrit), relevant for high-dimensional quantum key distribution and photonic encoding. The output dimension $M=4$ models dimension-altering channels due to ancillary states or noise. The number of Kraus operators $K=5$ represents moderate decoherence, consistent with noise models like amplitude damping and depolarizing channels \cite{Sim3}. These settings cover multi-qubit entanglement and high-dimensional encoding scenarios \cite{Sim1,Sim2}.

The input ensemble size is $P=N$, with $N$ linearly independent pure states forming a full-rank input density matrix, enabling maximal Holevo capacity \cite{Holevo1,Sim4}.

Initial probabilities $p_i$ follow a normalized Dirichlet distribution from uniform draws $\mathcal{U}(0,1)$. Initial input states $\mathbf{X}_i = \mathbf{x}_i \mathbf{x}_i^H$ are complex Gaussian vectors $\mathbf{x}_i \sim \mathcal{CN}(\mathbf{0}, \mathbf{I}_N)$, and then normalized as ${{\mathbf{x}}_i} \leftarrow {{\mathbf{x}}_i}/\left\| {{{\mathbf{x}}_i}} \right\|$ to guarantee the constraints in (\ref{con:1}).

Initial quantum channels are drawn from a normalized Rayleigh distribution. Specifically, ${{\mathbf{H}}_k}$ are a set of independently generated complex Gaussian matrices with entries drawn from $\mathcal{C}\mathcal{N}\left( {0,1} \right)$. To ensure that the resultant set defines a CPTP quantum channel, we normalize the operators as ${{\mathbf{H}}_k} \leftarrow {{\mathbf{H}}_k}{\left( {\sum\limits_{j = 1}^K {{\mathbf{H}}_j^H{{\mathbf{H}}_j}} } \right)^{ - 1/2}}$, so that the completeness condition in (\ref{channel:2}) is satisfied.

For simplicity, the step sizes $\alpha$, in (\ref{derivative:4}) is set to 0.3. Additionally, the maximum number of iterations is set to 100.

\begin{figure}[t]
\centering
\includegraphics[width=3.5in,height=2.7in]{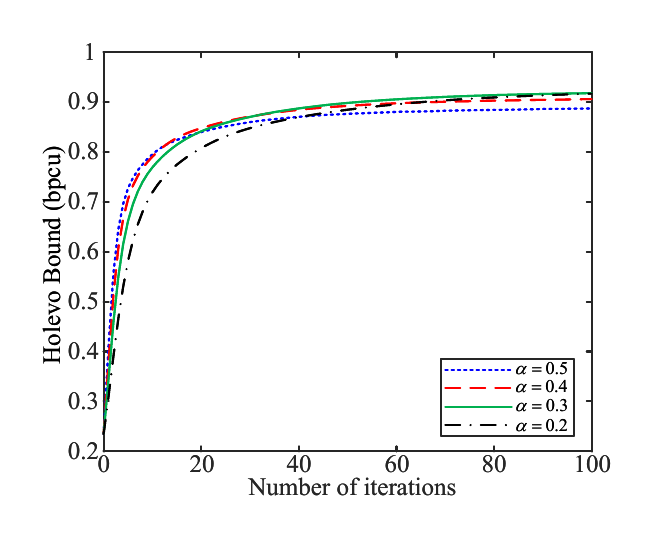}
\vspace{-3mm}
\caption{The convergence curves of the proposed AG algorithm with $N=3, M=4, K=5$.}
\label{fig_3}
\vspace{-0em}
\end{figure}

\subsection{Convergence}

Fig. \ref{fig_3} visualizes the convergence behavior of the proposed AG algorithm. Due to the efficient gradient formulations in (\ref{derivative:3}), the Holevo bound exhibits a monotonic increase with the number of iterations. For the configuration of $N=3, M=4$, and $K=5$, the algorithm achieves a Holevo bound exceeding 1.5 bpcu within 100 iterations. Moreover, variations in the step size from 0.2 to 0.5 have a marginal effect on the final converged value, with the resultant Holevo bound  being highest when the step size is set to 0.3.

%
%
%

\begin{figure}[t]
\centering
\includegraphics[width=3.5in,height=2.7in]{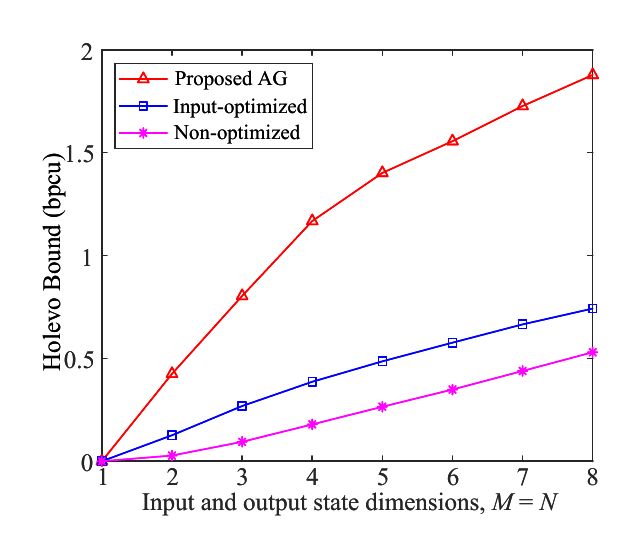}
\vspace{-3mm}
\caption{Comparison of Holevo bound across different schemes under varying input and output dimensions $N = M$.}
\label{fig_11}
\vspace{-0em}
\end{figure}

\begin{figure}[t]
\centering
\includegraphics[width=3.5in,height=2.7in]{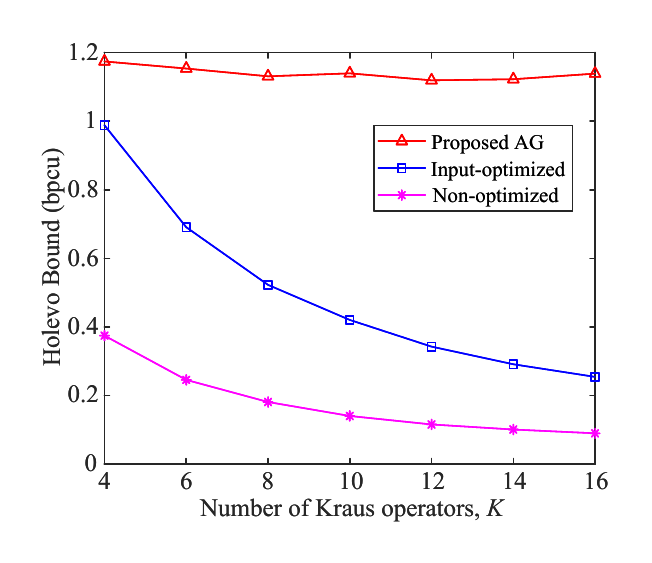}
\vspace{-3mm}
\caption{Comparison of Holevo bound across different schemes under varying number of Kraus operators $K$ with $N = M = 4$.}
\label{fig_12}
\vspace{-0em}
\end{figure}

\subsection{Comparison}

Fig. \ref{fig_11} shows the Holevo bound as a function of the input and output dimensions ($N = M$). As $N$ and $M$ increase,  all schemes exhibit improved capacity due to the expanded Hilbert spaces, which enhance the system's ability to encode and discriminate classical information. However, the growth for the proposed AG approach is sublinear, exhibiting diminishing returns at higher dimensions. This trend arises from reduced marginal gains in state distinguishability under fixed noise and power constraints, and the limited scalability of quantum channel fidelity with system size. These results underscore the importance of balanced dimensional design, since blindly increasing $N$ and $M$ yields limited capacity gains under practical constraints.

Fig. \ref{fig_12} quantifies the Holevo bound of different schemes under varying numbers of Kraus operators $K$, which characterize the level of noise and decoherence in the quantum channel. As $K$ increases, the Holevo capacities of the proposed AG for channel optimization exhibit a slight decline, whereas those of the input-only optimization and non-optimized scheme degrade more significantly. This is because an increased number of Kraus operators represents a more complex and noisy quantum environment, which reduces the channel's ability to preserve quantum information. The relatively smaller degradation observed in the proposed AG scheme indicates their enhanced robustness, owing to adaptive exploitation of the channel characteristics, while the lack of channel adaptation in the latter two schemes leaves them more vulnerable to noise-induced capacity loss.

\emph{In a nutshell, the proposed AG approach outperforms conventional input optimization and non-optimized schemes. This superior performance results from the exploitation of channel adaptation, facilitating a more effective exploitation of the channel's degrees of freedom.}

\section{Conclusion}

We proposed a GA algorithm to maximize the Holevo bound of a quantum system by optimizing the quantum channel given fixed quantum inputs, while preserving the CPTP maps through projection operations.
A detailed complexity analysis demonstrated the algorithm's scalability with respect to system parameters such as dimension and channel rank. Simulation results confirm that the proposed method significantly outperforms input-only baselines, especially for high-rank quantum channels. Overall, the proposed method offers a practical and theoretically grounded tool for optimizing classical information throughput in programmable quantum communication systems.

\section*{Acknowledgments}
This work was supported by the MoE AcRF Tier 1 Thematic Grant RT12/23 023780-00001. The financial support of the following Engineering and Physical Sciences Research Council (EPSRC) projects is also gratefully acknowledged: Platform for Driving Ultimate Connectivity (TITAN) under Grant EP/Y037243/1 and EP/X04047X/1; Robust and Reliable Quantum Computing (RoaRQ, EP/W032635/1); India-UK Intelligent Spectrum Innovation ICON UKRI-1859; PerCom (EP/X012301/1); EP/X01228X/1; EP/Y037243/1.

\setcounter{subsubsection}{0}

\ifCLASSOPTIONcaptionsoff
  \newpage
\fi

\end{document}